# Fabrication of a microresonator-fiber assembly maintaining a high-quality factor by $CO_2$ laser welding

Zhiwei Fang,[1,2] Jintian Lin,[2] Min Wang,[2,3] Zhengming Liu,[1,2] Jinping Yao,[2] Lingling Qiao,[2] and Ya Cheng[1,2,*]

[1]*School of Physical Science and Technology, ShanghaiTech University, Shanghai 200031, China*
[2]*State Key Laboratory of High Field Laser Physics, Shanghai Institute of Optics and Fine Mechanics, Chinese Academy of Sciences, P.O. Box 800-211, Shanghai 201800, China*
[3]*University of Chinese Academy of Sciences, Beijing 100049, China*
[*]*ya.cheng@siom.ac.cn*

**Abstract:** We demonstrate fabrication of a microtoroid resonator of a high-quality (high-Q) factor using femtosecond laser three-dimensional (3D) micromachining. A fiber taper is reliably assembled to the microtoroid using $CO_2$ laser welding. Specifically, we achieve a high-Q-factor of $2.12 \times 10^6$ in the microresonator-fiber assembly by optimizing the contact position between the fiber taper and the microtoroid.

## 1. Introduction

A whispering-gallery-mode (WGM) microresonator of a high quality (high-Q) factor can be of great use for nonlinear optics and high-sensitivity sensing applications thanks to its capability of efficient confinement of light in a small volume by total internal reflection [1-7]. To couple light into or out of the high-Q microresonators, a fiber taper coupling technique is frequently used because of its high coupling efficiency and ease of operation [8]. Typically, to obtain the highest Q-factor with the fiber taper coupling, a critical coupling condition should be satisfied which relies on precise control of the distance between the microresonator and the fiber taper. The distance of the critical coupling condition is determined by the decay length of the evanescent field of the WGMs in the microresonator, which is usually in the range of a few hundred of nanometers. Although the critical coupling can be readily realized with nano-positioning stages, the coupled microresonator-fiber system usually cannot be easily translated between different locations because of the challenge in the maintaining of the sub-micron distance between the microresonator and the fiber taper which are physically separated from each other. It is also difficult to mount the 3D nano-positioning stage in an optical cryostat for investigating cavity quantum electrodynamics (cavity-QED) [9]. Particularly, the difficulty can be more severe for the future commercialization of the microdevices (such as sensors, microresonator lasers, etc.) incorporated with high-Q microresonators.

Previously, we have demonstrated assembly of the fiber taper to a microresonator by directly welding the fiber taper onto the sidewall of the microtoroid using $CO_2$ laser annealing [10]. The integrated microresonator-fiber assembly (MFA) showed a Q-factor of $3.21\times10^5$ as measured in air, which is approximately one order of magnitude lower than the Q-factor achieved with the same microresonator in the critical coupling condition. The reduction in the Q-factor is mainly caused by the direct contact of the fiber taper to the sidewall of the microtoriod, making the operation of MFA in the over coupling regime. Here, we systematically investigate the dependence of the Q-factor of the MFA on the contact position between the fiber taper and the microtoroid. It is found that when the contact position is optimized, the Q-factor of the MFA can be improved by almost one order of magnitude, which reaches $2.12\times10^6$.

## 2. Experiment

In this work, commercially available fused silica glass substrates (UV grade fused silica JGS1 whose six surfaces are polished to optical grade) with a thickness of 2 mm are used. In our previous works, the fabrication technology of high-Q microtoroid resonators based on femtosecond laser direct writing has been established. The details can be found in [10,11]. Here, we only briefly describe the fabrication process, which mainly includes three steps: (1) femtosecond laser exposure followed by selective wet etch of the irradiated areas to create a microdisk structure; (2) selective reflow of the silica disk by $CO_2$ laser irradiation to improve the Q-factor; and (3) assembly of the fiber taper with the microresonator by $CO_2$ laser welding. The process is schematically illustrated in Fig. 1.

The experimental conditions are provided as follows. The femtosecond laser system used for fabrication of the microtoroids consists of a Ti: sapphire oscillator (Coherent, Inc.) and a regenerative amplifier, which emits 800 nm, ~40 fs pluses with a maximum pulse energy of ~4.5 mJ at 1-kHz repetition rate. Power adjustment was realized using a variable optical attenuator. The glass samples could be arbitrarily translated in 3D space at a resolution of 1 μm using a PC-controlled XYZ stage combined with a nano-positioning stage. In the femtosecond laser direct writing, a 100× objective lens with a numerical aperture (NA) of 0.80 was used to focus the beam down to a ~1 μm-dia. focal spot, and the average laser power measured before the objective was ~0.05 mW. After the laser direct writing, the sample was subjected to a ~40 min bath in a solution of 5% HF diluted with water to form an on-chip microdisk. To achieve the high-Q-factor, we smooth the microtoroid surface by surface reflow using $CO_2$ laser (Synrad Firestar V30) annealing. The fabricated microtoroid has a diameter of ~70 μm and a thickness of ~8 μm.

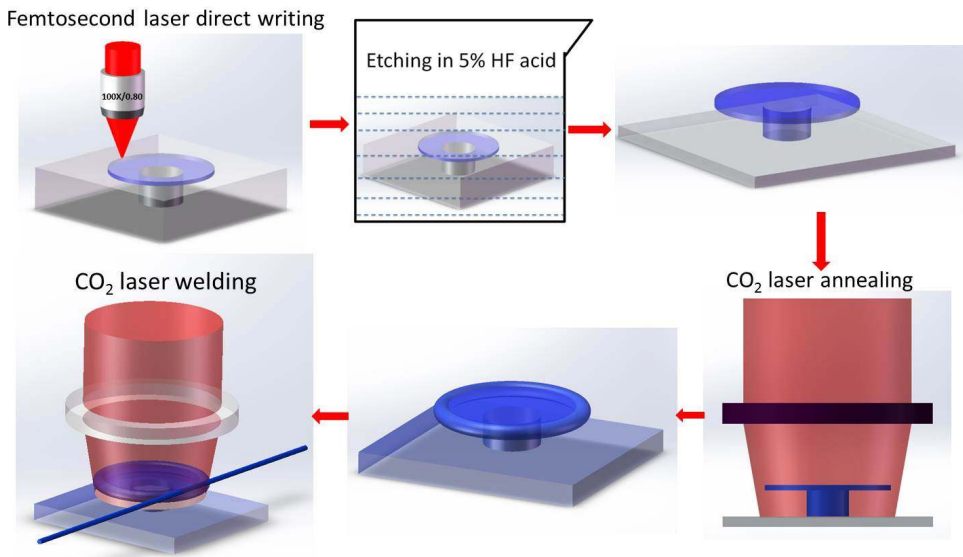

Fig. 1. Schematic illustration of fabrication of a microresonator using femtosecond laser direct writing onto which a fiber taper is assembled by $CO_2$ laser welding.

To measure the Q-factor of the microtoroid, resonance spectra were measured via the optical fiber taper coupling method [8]. A swept-wavelength external-cavity tunable diode Laser (Model: 6528-LN, New Focus) operating at 1550 nm range with linewidth of 10 MHz and a swept spectrometer (Model: 4650, dBm Optics Inc.) were used to measure the transmission spectrum from the fiber taper at a resolution 0.2 pm. The microtoroid was fixed on a three-axis nano-positioning stage with a resolution of 50-nm in XYZ direction. We used dual charge coupled device (CCD) cameras to simultaneously image the microtoroid and the fiber taper from the side and top.

To assemble the fiber taper with the microtoroid resonator, the fiber taper was physically attached to the sidewall of the microtoroid and then the microtoroid was irradiated with the $CO_2$ laser beam for welding the fiber taper onto the microtoroid. To ensure a high Q-factor, the key is to optimize the contact position between the fiber taper and the microtoroid, which determines the overlapping between the mode in the microtoroid and that in the fiber taper. The process of the optimization is discussed in detail below.

### 3. Results and discussion

For optimizing the contact position between the fiber taper and the microtoroid, first, the fiber was brought close to the microtoroid to achieve the critical coupling. Figure 2(a) shows the geometric configuration of the critical coupling between the fiber taper and the microtoroid. In this case, there is a sub-wavelength gap between the fiber and the microtoroid, and the coupling strength can be precisely controlled by tuning the width of the gap. The transmission spectrum (FSR) obtained is shown in Fig. 2(b), we observe a free spectral range of 7.49 nm which is consistent with the theoretically calculated value of a 70-μm-diameter microtoroid. The calculation was performed using the following expression [11]

$$\Delta\lambda_{FSR} \approx \frac{\lambda_0^2}{n\pi D} ,  \qquad (1)$$

where $\lambda_0$ is the wavelength in vacuum, $D$ the diameter of the microresonator, and $n$ the refractive index of the fused silica. For $D\approx 70$ μm and $n\approx 1.445$, the theoretically predicted value of $\Delta\lambda_{FSR}$ is about 7.47 nm at 1548.01 nm. The Q-factor obtained in near critical coupling condition was calculated to be $2.55\times 10^6$ based on a Lorentz fitting, as evidenced by the red curve in Fig. 2(c). The similar Q-factor has also been demonstrated in previous experiments of fabricating high-Q microresonators using the femtosecond laser [10-13].

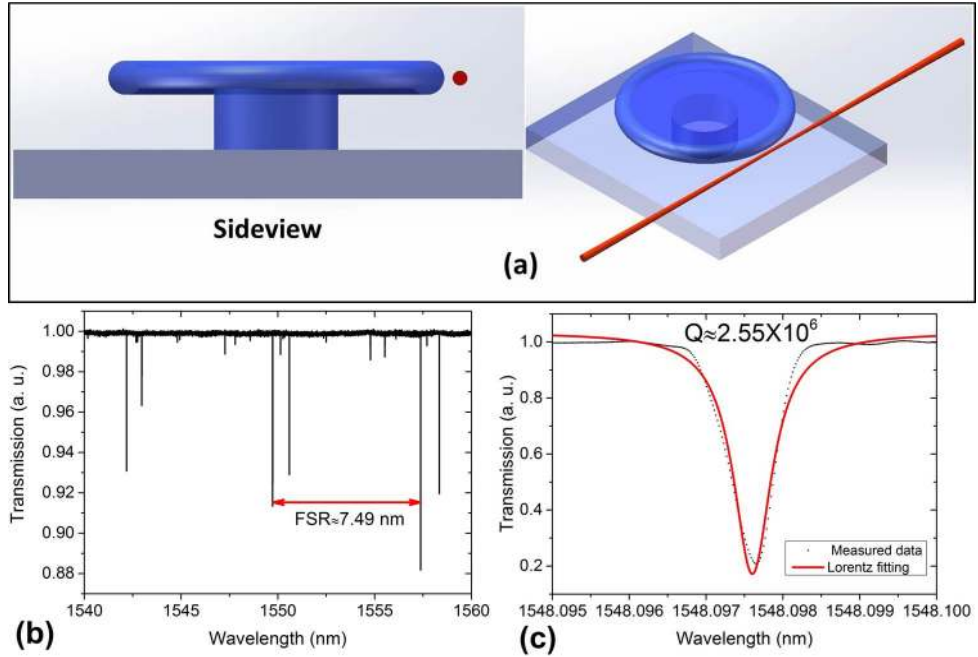

Fig. 2. (a) The geometric configuration of the fiber taper and the microtoroid in the critical coupling condition (Left: sideview; right: top view). (b) The transmission spectrum of the microtoroid. (c) The Lorentz fitting (red curve) showing a Q-factor of $2.55\times 10^6$.

Next, we translated the fiber taper horizontally toward the microtoroid until a physical contact between the fiber and the microtoroid was achieved, as schematically illustrated in Fig. 3(a). We then recorded the transmission spectrum again which is shown in Fig. 3(b). Due to the over coupling between the fiber taper and the microtoroid, we observed a significant drop in the Q-factor as indicated by the Lorentz fitting curve in Fig. 3(c). The Q-factor was measured to be $2.64\times10^5$, which is consistent with that in our previous work [10]. In addition, we observe that the transmission spectrum appears chaotic with densely distributed peaks, indicating the excitation of high-order modes in the microtoroid.

To promote the Q-factor of the microtoroid, it is critical to optimize the coupling strength. This can be achieved by adjusting the contact position between the fiber taper and the microtoroid. Therefore, we first slightly shifted the fiber taper in the upward direction and in the lateral direction to reduce the overlapping between the mode in the fiber and that in the microtoroid, as illustrated in Fig. 4(a). In this case, due to the taper of the fiber, the diameter of the fiber increases to ~3 μm at the contact position. In addition, in the upper area away from the equator of the microtoroid, the field strength becomes weaker than that at the equator. Both of these lead to a reduction of the coupling strength, which gives rise to higher Q-factors similar to that obtained in the critical coupling condition. After optimization of the contact position, we measured the transmission spectrum of the microtoroid again which is shown in Fig. 4(b). Remarkably, the measured Q-factor reaches $2.24\times10^6$ as evidenced by Fig. 4(c). The experimental results are qualitatively consistent with the prediction of previous theoretical simulation [14]. Besides, the transmission spectrum in Fig. 4(b) is much cleaner than that in Fig. 3(b). The Q-factor is on the same level of that achievable in the critical coupling condition, however, the physical contact between the fiber and the microtoroid provides the possibility to produce a fully integrated, portable MFA.

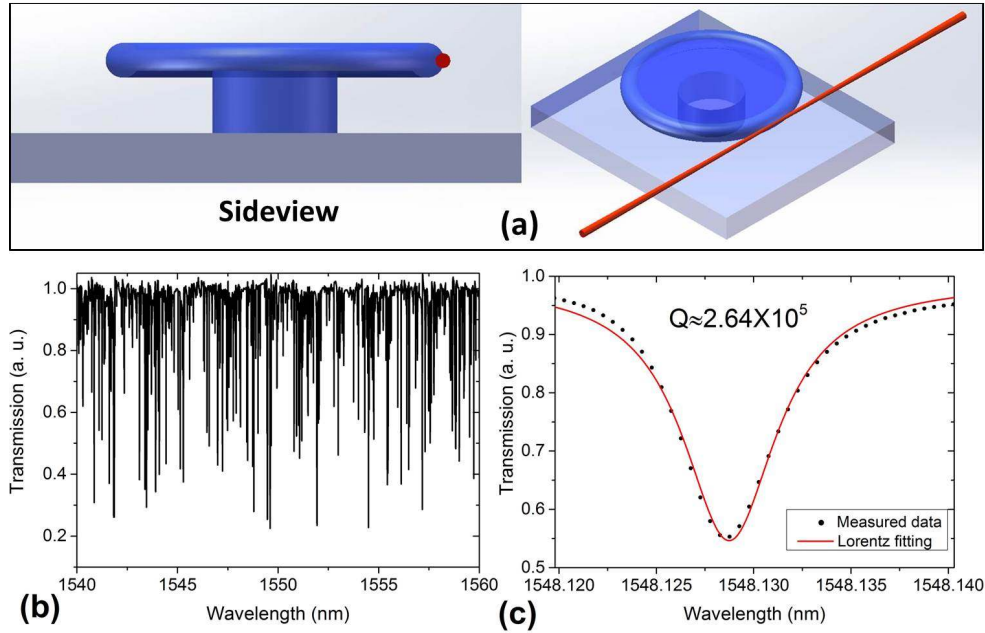

Fig. 3. (a) The geometric configuration of the fiber in direct contact with the microtoroid. (b) The transmission spectrum of the microtoroid. (c) The Lorentz fitting (red curve) showing a Q-factor of $2.64\times10^5$.

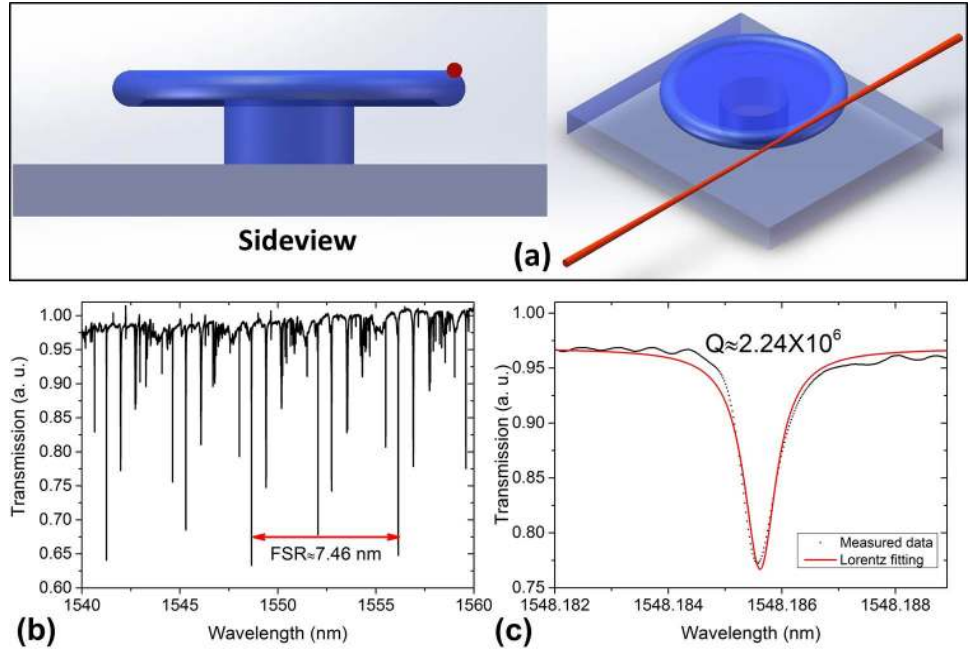

Fig. 4. (a) The geometric configuration of the fiber taper in direct contact with the microtoroid after the optimization of the contact position. (b) The transmission spectrum of the microtoroid. (c) The Lorentz fitting (red curve) showing a Q-factor of $2.24 \times 10^6$.

As the last step, we performed the $CO_2$ laser welding to produce the fully integrated MFA. The recorded transmission spectrum of the MFA after the $CO_2$ laser welding is shown in Fig. 5(a). The Q-factor was determined again by the Lorentz fitting as shown by the red curve in Fig. 5(b), which reaches $2.12 \times 10^6$. In addition, as is shown in Fig. 5(c), it was also observed that the resonant peak around 1548 nm slightly shifted to the blue side by a $\Delta\lambda \approx$ 0.0015 nm after the $CO_2$ laser welding. This can be attributed to the slight deformation of the microtoroid resonator induced by the $CO_2$ laser annealing. Since there are several approaches to tune the resonance peaks of the microtoroid resonator, the spectral shift induced by the $CO_2$ laser welding can be readily compensated in practical applications.

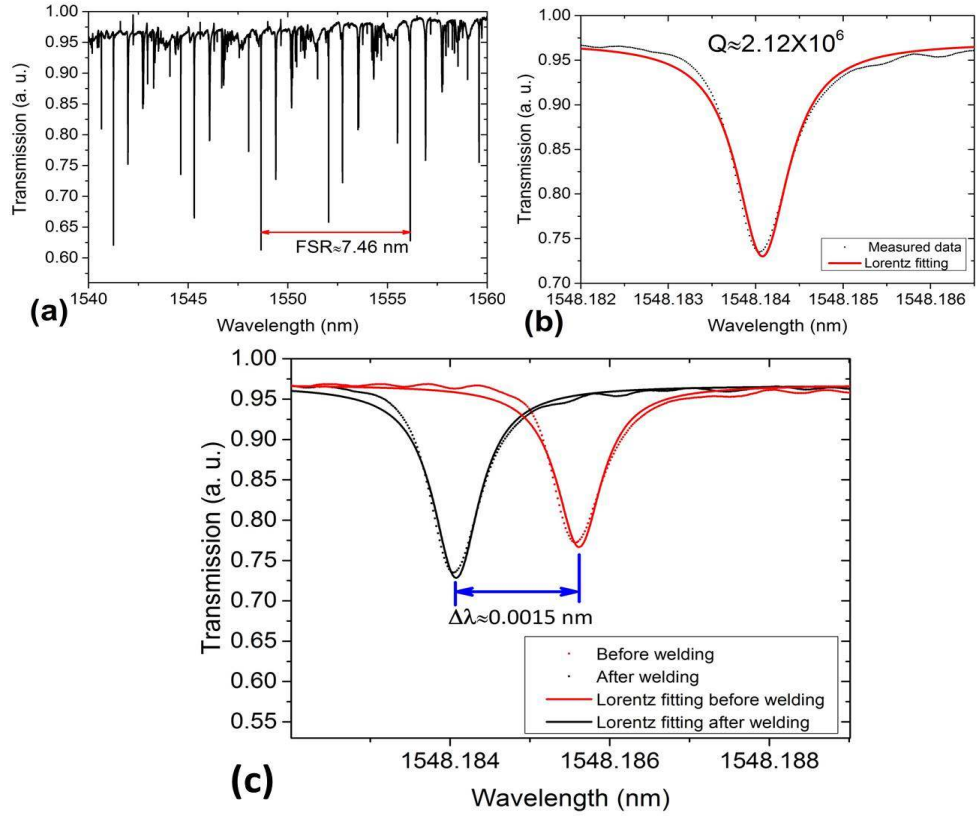

Fig. 5. (a) The tansmission spectrum of the fabricated MFA. (b) The Lorentz fitting (red curve) showing a Q-factor of $2.12\times10^6$. (c) Shifting of the resonant peak after the $CO_2$ laser welding.

## 4. Conclusion

To conclude, we have developed a technique to realize integrated MFA with a Q-factor as high as $2.12\times10^6$. Our technique based on $CO_2$ laser welding is simple and easy to implement. The key to maintain the high-Q-factor is to adjust the contact position between the microresonator and the fiber taper, by which the mode overlap can be optimized. With the high bonding strength between the fiber and the microtoroid, the miniaturized and portable high-Q MFA will find potential applications in various fields ranging from biological and chemical sensing [15,16] to nonlinear optics [17,18] and cavity-QED [1,9].

**Acknowledgments**

The authors would like to thank Prof. Tao Lu of University of Victoria for fruitful discussion.